\documentstyle[12pt,epsfig]{article}
\begin{document}
\noindent
\begin{center}
{\Large {\bf  Redshift Dependence of CMB Temperature \\in BSBM $\alpha$-Varying Theories }}\\
\vspace{2cm}
 ${\bf Yousef~Bisabr}$\footnote{e-mail:~y-bisabr@sru.ac.ir.}\\
\vspace{0.5cm} {\small{Department of Physics, Shahid Rajaee Teacher
Training University,
Lavizan, Tehran 16788, Iran}}\\
\end{center}
\vspace{1cm}
\begin{abstract}
We have investigated some cosmological consequences of variation of the fine structure constant, $\alpha$, in the context of
the Bekenstein-Sandvik-Barrow-Magueijo (BSBM) model. In this context, such a variation is attributed to electric charge by
letting the latter take on the value of a real scalar field, $\phi$,  which varies generally in space and time. We consider the dynamics of $\phi$ in the case that it is allowed to have an exponential potential. We have found solutions for evolutions of $\phi$ and $\alpha$ in the radiation and the matter dominated eras. By employing observational bounds on the temporal variation of $\alpha$, we constrain the model parameters. We then investigate the impact of $\alpha$-variation on redshift dependence of CMB temperature. It is shown that there is a deviation from the standard evolution of the temperature which contains two different contributions, one is important at high redshift eras and the other is mostly effective at low redshift eras.
\end{abstract}
~~~~~~~~~~~~~~~~~~~~~~~~~~~~~~~~~~~~~~~~~~~~~~~~~~~~~~~~~~~~~~~~~~~~~~~~~~~~~~~~~~~~~~~~~~~~~~~~~~~~~
\section{Introduction}
Many years after Dirac's proposal \cite{di} that fundamental physical constants may vary with time and/or
space, there is now some theoretical frameworks which predict such variations. For instance,
one of the most interesting low energy features of string theory is the presence of
massless scalar fields (dilatons or moduli fields) whose vacuum expectation values define the size
of the effective coupling constants. Independent of this framework, there is a large class of dark energy and
quintessence models which invoke scalar fields with wavelengths comparable to the size of the
Universe. These dilatons or quintessence fields may then interact with matter and radiation so that changes of its
background value induce variation of coupling and fundamental constants \cite{ca}.\\
There is an inherent ambiguity concerning measurement of changes of a
dimensional quantity. In general, the value of a dimensional quantity not only may change
in a given unit system but it may also change due to changes of the unit system by which
the quantity is measured. There is not however any direct way to distinguish between these
two types of changes. Thus, it is only meaningful to discuss about dimensionless combinations of fundamental constants. Central attention is devoted to the fine structure constant ($\alpha=e^2/\hbar c$) as the dimensionless coupling constant of electromagnetic interactions. Variations of $\alpha$ may be theoretically attributed to each of the physical quantities $e$, $\hbar$ or $c$. Although theories based on variation of each of theses quantities are physically equivalent, they may need quite different mathematical structures. There are two main approaches that model possible variations of $\alpha$. In one approach, variation of $\alpha$ is attributed to varying speed of light \cite{vsl} which it conflicts with Lorentz invariance. In another approach the electric charge is taken to be varied with the cosmic time. This should be constrained by experiments verifying charge conservation \cite{lan}. The latter approach was developed by Bekenstein \cite{bek} who generalized the Maxwell's equations to incorporate possible variation of electron charge. He provided a varying-$e$ theory preserving local gauge invariance and Lorentz invariance. Later, the theory was extended to include gravity and its application to a cosmological setting was proposed \cite{bsbm}. This extension is referred to as BSBM theory. In this theory, variations of $e$ are encoded in a scalar field whose interaction with electromagnetic sector induces variation of $\alpha$. Such an interaction has different consequences which one of them is modification of redshift dependence of temperature of the cosmic microwave background (CMB) radiation in the Universe. In fact, photons of the radiation left over after decoupling from matter travel freely through space and loose their energy due to expansion of the Universe. The standard picture is that the temperature of the CMB increases linearly with redshift, namely that
\begin{equation}
T(z)=T_0(1+z)
\label{0-0}\end{equation}
with $T_0=2.72548\pm 0.00057K$ being the temperature at $z=0$ \cite{t}. However, there are a number of physical
processes that could affect this temperature-redshift relation. Different cosmological models that consider non-conservation of photon numbers are the most important processes of this kind. For instance, it is shown \cite{l} that decay of vacuum energy density provides a source term for radiation and affects the standard temperature evolution law. In BSBM model there is an energy transfer between radiation and the dynamical scalar field driving evolution of $\alpha$. Due to this energy exchange between the two components temperature-redshift law deviates from the standard picture. One usually parametrizes such a deviation as \cite{lima}
\begin{equation}
T(z)=T_0(1+z)^{1-\beta}
\label{0-1}\end{equation}
where $\beta$ is a dimensionless parameter. Recent developments of techniques used to measure temperature of CMB at $z\sim1$ constrain the parameter $\beta$ to be $0.004\pm 0.016$ \cite{non}.
In this work we will show that the expression (\ref{0-1}) is only an approximate temperature-redshift law which holds only at high redshift eras. At low redshift eras, modification of $T(z)$ from the standard one will be slightly different from (\ref{0-1}).\\
This work is organized as follows: In section 2, we review the BSBM model in which the scalar field driving variation of $\alpha$ has an exponential potential.
Due to interaction with the radiation sector, the balance equation of the radiation has two source terms ($Q_1$ and $Q_2$ in the following). We find solutions of the field equations in the radiation and matter dominated eras. In section 3, we investigate the impact of variation of $\alpha$ on the redshift dependence of the CMB temperature. We argue that although $Q_2$ is not important in dynamics of $\alpha$, it can affect the evolution of temperature, specifically, at late times. We will show that $T(z)$ is given by (\ref{0-1}) in the radiation-dominated era while it follows a different modification in the matter-dominated era. In section 4, we reach our conclusions.

~~~~~~~~~~~~~~~~~~~~~~~~~~~~~~~~~~~~~~~~~~~~~~~~~~~~~~~~~~~~~~~~~~~~~~~~~~~~~~~~~~~~~~~~~~~~~~~~~~~~~~~~~~~~~~~~~~~~~~~~~~~~~~~~~
\section{The BSBM model: Revisited}
In BSBM model, variation of $\alpha$ is attributed to changes of $e$ while $c$ and $\hbar$ are taken to be constants \cite{bek} \cite{bsbm}. This is done by letting $e$ take on the value of a real scalar field which may generally vary in
space and time, thus
\begin{equation}
e_0\rightarrow e=e_0 ~\varepsilon(x^\mu)
\label{1-1}\end{equation}
where $\varepsilon$ is a dimensionless scalar
field and $e_0$ is a constant denoting the present value of $e$. This is consistent with the principles of local gauge invariance and causality
\cite{bek} and should be constrained by
charge conservation \cite{lan}. Since the field $\varepsilon$ couples with the electromagnetic gauge field as
$a_{\mu}\equiv\varepsilon A_{\mu}$, one should generalize the usual gauge transformation $A_{\mu}\rightarrow
A_{\mu}+\chi_{,\mu}$ to
\begin{equation}
a_{\mu}\rightarrow a_{\mu}+\chi_{,\mu}
\label{1-2}\end{equation}
where $\chi$ is an arbitrary function. The gauge-invariant electromagnetic field tensor is then
\begin{equation}
f_{\mu\nu}=\varepsilon F_{\mu\nu}=\partial_{\mu}a_{\nu}-\partial_{\nu}a_{\mu}
\label{1-3}\end{equation}
It reduces to the usual form when $\varepsilon$ takes a constant value. The electromagnetic Lagrangian is $
L_{em}=-\frac{1}{4}f^{\mu\nu} f_{\mu\nu}$ where the dynamics of $\varepsilon$ are controlled by
$L_{\varepsilon}= -\frac{1}{2}\frac{\omega}{\varepsilon^2}(\varepsilon_{,\mu}\varepsilon^{,\mu})$ with
$\omega$ being a coupling constant\footnote{The BSBM model can be generalized to include the case that $\omega$ appears as a coupling function $\omega(\phi)$ \cite{g}.} defined by $\omega=\frac{\hbar
c}{l^2}$. The quantity $l$ is the characteristic length scale of
the theory introduced for dimensional reasons. Note that $\varepsilon$ is dimensionless and we need $l$ in the Lagrangion to have a correct dimension. There is an upper bound\footnote{There is also a lower bound given by the Planck length \cite{bek}. } on this characteristic length  $l<10^{-15}cm$ coming from the observed scale invariance of electrodynamics \cite{bek}.\\
In a cosmological setting, variation of
$\alpha$ may be investigated by the following action functional \cite{bsbm}
\begin{equation}
S_{total}=\int ~d^4x~\sqrt{-g}~[L_g+L_{m}+L_{\phi}+L_{em}e^{-2\phi}]
\label{1-4}\end{equation}
where $L_g=\frac{1}{16\pi G}R$ is the Einstein-Hilbert term and $L_{m}$ is Lagrangian of the matter system.
The earlier analysis of the BSBM model considered the situation in which the
scalar field has no self-interaction potential and $L_{\phi}=-\frac{\omega}{2}g^{\mu\nu}\partial_{\mu}\phi
\partial_{\nu}\phi$ with $\phi\equiv\ln\varepsilon$. In a more general case, the scalar field driving
variations of $\alpha$ may have a non-zero
potential \cite{pot}. We will consider the case that $L_{\phi}=-\frac{\omega}{2}g^{\mu\nu}\partial_{\mu}\phi
\partial_{\nu}\phi-V(\phi)$ with $V(\phi)$ being a potential function.\\
The theory described by the above action generalizes
Bekenstein's approach by including variations of $\varepsilon$ on the gravitational dynamics
of the expanding Universe. Taking (\ref{1-4}) as a part of the low-energy effective action of string theory,
the scalar field $\phi$ plays the role of a dilaton or a moduli field. In principle, these scalar fields can
couple to all parts of the matter system and induce variations of all fundamental and coupling constants. In
the BSBM model, described by the action (\ref{1-4}), the field $\phi$ is assumed to interact only with
electromagnetic energy. In this work, we also restrict ourself to this case but relax this assumption and consider a more general case elsewhere.\\
Variation of (\ref{1-4}) with respect to $g^{\mu\nu}$ gives gravitational equations
\begin{equation}
G_{\mu\nu}=8\pi G(T_{\mu\nu}^{m}+T_{\mu\nu}^{\phi}+ e^{-2\phi} T_{\mu\nu}^{em})
\label{1-5}\end{equation}
and with respect to $\phi$ gives
\begin{equation}
\Box \phi-\frac{1}{\omega}V'(\phi)=\frac{2}{\omega}e^{-2\phi} L_{em}
\label{1-6}\end{equation}
where $V'(\phi)\equiv \frac{dV(\phi)}{d\phi}$. The RHS of (\ref{1-6}) vanishes for pure radiation. We recall that electric and magnetic fields have similar contributions to electromagnetic energy and
$L_{em}=(E^2-B^2)/2=0$. Thus only can non-relativistic matter be a non-vanishing source of (\ref{1-6}). There is, however, a non-zero contribution of electromagnetic energy to the mass of particles
which can be parametrized by $\zeta\equiv L_{em}/\rho_m$ with $\rho_m$ being energy density of
matter \cite{bek} \cite{bsbm}.
In a cosmological context, $\zeta$ is highly a constant \cite{bek} which depends on both baryonic matter and dark matter. Thus it can take both positive and negative values depending on which of Coulomb or magnetostatic energy dominates the dark matter of the Universe \cite{bsbm}.
Due to strong uncertainties in the nature of dark matter and its constituents, there is no a firm bound on $\zeta$. Following the earlier analyses, we consider the case that $\frac{|\zeta|}{\omega}\sim 0.01\%$.\\ By solving the equation (\ref{1-6}), the dynamics of $\alpha$ is then given by $\alpha=\alpha_0 e^{2\phi}$ with $\alpha_0\equiv\alpha(z=0)$.
Applying the Bianchi identities to (\ref{1-5}) gives
\begin{equation}
\nabla^{\mu} T_{\mu\nu}^{m}=0
\label{1-7}\end{equation}
\begin{equation}
\nabla^{\mu}(e^{-2\phi} T_{\mu\nu}^{em})=-\nabla^{\mu}T_{\mu\nu}^{\phi}
\label{1-7}\end{equation}
which indicates that $T_{\mu\nu}^{em}$ is not conserved due to interaction with $\phi$. For a spatially flat
Friedmann-Robertson-Walker metric
\begin{equation}
ds^2=-dt^2+a^2(t)(dx^2+dy^2+dz^2)
\end{equation}
with $a(t)$ being the scale factor, the field equations become\footnote{We work in units in which $\hbar=c=1$.}
\begin{equation}
3H^2=8\pi G[(1+|\zeta| e^{-2\phi})\rho_m+e^{-2\phi}\rho_r+\rho_{\phi}]
\label{1-11}\end{equation}
\begin{equation}
\dot{\rho}_m+3H\rho_m=0
\label{1-8}\end{equation}
\begin{equation}
\dot{\rho}_r+4H\rho_r=Q_1+Q_2 e^{-2\phi}
\label{1-9}\end{equation}
\begin{equation}
\dot{\rho}_{\phi}+3H(\omega_{\phi}+1)\rho_{\phi}=-Q_2 e^{-2\phi}
\label{1-10}\end{equation}
where $\rho_{\phi}=\frac{1}{2}\omega\dot{\phi}^2+V(\phi)$, $p_{\phi}=\frac{1}{2}\omega\dot{\phi}^2-V(\phi)$, $H=\frac{\dot{a}}{a}$ and $\omega_{\phi}=\frac{p_{\phi}}{\rho_{\phi}}$. The RHS of the balance equations of radiation and $\phi$ are not zero due to existence of the source terms $Q_1$ and $Q_2$ which are defined by
\begin{equation}
Q_1=2\dot{\phi}\rho_r
\label{1-10a}\end{equation}
\begin{equation}
Q_2=2|\zeta|\dot{\phi}\rho_m
\label{1-10b}\end{equation}
Here $\rho_m$ and $\rho_r$ are matter and radiation energy densities,
respectively. By defining $\bar{\rho}_r=e^{-2\phi}\rho_r$, the equation (\ref{1-9}) takes the form
\begin{equation}
\dot{\bar{\rho}}_r+4H\bar{\rho}_r=Q_2 e^{-4\phi}
\label{1-9ab}\end{equation}
The matter system is uncoupled and thus the equation (\ref{1-8}) is satisfied by the standard solution $\rho_m=\rho_{0m}a^{-3}$ where $\rho_{0m}$ is an integration constant. The remaining equations (\ref{1-11}), (\ref{1-9}) and (\ref{1-10}) are going to be solved in the limit $Q_2=0$. We also take $V(\phi)=V_0 e^{-\lambda\phi}$ with $V_0$ and $\lambda$ being constants.
Besides smallness of $|\zeta|$, there is another factor that weakens the effect of $Q_2$. If $\alpha$ changes, its rate of change which is given by $\dot{\phi}$ must be sufficiently small as inferred by observations. The appearance of two small coefficients in $Q_2$ supports its ignorance.
In the next section, however, we will see that in spite of the negligible role of $Q_2$ in dynamics of $\alpha$, it affects the temperature-redshift relation of CMB.\\Discarding $Q_2$ from (\ref{1-9ab}) yields $\bar{\rho}_r=\bar{\rho}_{0r}a^{-4}$ with $\bar{\rho}_{0r}$ being an integration constant. Thus $\bar{\rho}_r$ has the same scaling behavior as radiation energy density in the standard cosmology with a constant $\alpha$. It should be noted that $\bar{\rho}_r$ and $\rho_r$ have different scalings because even though $Q_2=0$, there is still the source term $Q_1$ in (\ref{1-9}) which causes photon non-conservation. These energy densities have been the origin of a controversy concerning which of the two is the suitable definition of radiation energy density in BSBM theory. In the viewpoint that takes $\rho_r$ as the correct definition, the photon number is not conserved due to the source term $Q_1$ and the temperature-redshift relation becomes modifed \cite{non}. There is another viewpoint which takes $\bar{\rho}_r$ as a suitable definition \cite{ma}. In this case photon number is conserved and the temperature-redshift relation is not modified.\\
We seek solutions of (\ref{1-11}) and (\ref{1-10}) for which the scale factor behaves as $a(t)=a_0 t^{n}$ with $n=\frac{1}{2}, \frac{2}{3}$ in radiation and matter dominated eras. Let us check
 \begin{equation}
\phi=\phi_0+\frac{2}{\lambda} \ln(\frac{t}{t_0})
\label{1-12}\end{equation}
as the solution of (\ref{1-10}) which gives an algebraic relation
\begin{equation}
V_0 t_0^2 e^{-\lambda\phi_0}=\frac{2\omega(3n-1)}{\lambda^2}
\label{1-166}\end{equation}
We also need to check the Friedmann equation. Evaluating the terms and putting into (\ref{1-11}) and noting the fact that
$\bar{\rho}_r >>\rho_m$ for radiation domination, yields
\begin{equation}
\frac{3}{4}=8\pi G\{\bar{\rho}_{0r}a_0^{-4}+(\frac{2}{\lambda^2}+V_0 e^{-\lambda\phi_0})t_0^2\}
\label{1-16}\end{equation}
For matter domination $\bar{\rho}_r <<\rho_m$, one obtains
\begin{equation}
\frac{4}{3}=8\pi G\{\rho_{0m}a_0^{-3}+(\frac{2}{\lambda^2}+V_0 e^{-\lambda\phi_0})t_0^2\}
\label{1-16}\end{equation}
Thus the expression (\ref{1-12}) gives the scaling solution of $\phi$ and also determine time evolution of the fine structure constant\footnote{A phenomenological description of variation of $\alpha$ at the recombination era is given by \cite{re}.} as $\alpha\propto t^{\frac{4}{\lambda}}$ and then $\frac{\dot{\alpha}}{\alpha}=\frac{4}{\lambda}t^{-1}$. There is a stringent bound on any temporal drift in the value of $\alpha$ \cite{al}
\begin{equation}
|\frac{\dot{\alpha}}{\alpha}|=3.3\pm3.0\times 10^{-16}~yr^{-1}
\label{1-16a}\end{equation}
which constrains the parameter $\lambda$ as $\lambda\sim 10^{6}$. This parameter gives slope of the change of $\phi$, $\dot{\phi}=\frac{2}{\lambda}t^{-1}$, so that the larger it is the smaller the rate of change of $\phi$ and $\alpha$ to be. Thus we expect $\phi$ and then $\alpha$ to change very slowly during the course of evolution.

××××××××××××××××~×××××××××××××××××××××××××××××××××××××××××××××××××××××××××××××××××××××××××××××××××××××××××÷
\section{Modified temperature-redshift law}
Due to the interaction of $\phi$ with the electromagnetic sector, the evolution equations for the radiation energy density $\rho_{r}$ and number density $n_{r}$ are given by (\ref{1-9}) and
\begin{equation}
\dot{n}_r+3H n_r=\Psi
\label{1-17}\end{equation}
where $\Psi$ is a source term which is related to $Q_1+Q_2 e^{-2\phi}$. The interaction will in general distort the behavior
of the
radiation fluid, and in particular the photon temperature-redshift relation. We restrict ourselves to the
observationally relevant case of adiabatic evolution, that is when the entropy per photon remains constant. In
this case, one can write \cite{lima}
\begin{equation}
\frac{\dot{T}}{T}+H=\frac{\Psi}{3n_{r}}=\frac{Q_1+Q_2 e^{-2\phi}}{4\rho_{r}}
\label{1-20}\end{equation}
When $\Psi=0$, the standard evolution $Ta=T_0 a_0$ or, equivalently, the expression (\ref{0-0}) is recovered.
That the RHS of (\ref{1-20}) is non-vanishing implies that evolution of the temperature deviates from the standard one, such that
\begin{equation}
T(z)=T_0(1+z)\tau(z)
\label{1-21}\end{equation}
where $\tau(z)$ is a modification function. It is given by
\begin{equation}
\frac{d\tau}{\tau}=\frac{Q_1+Q_2 e^{-2\phi}}{4\rho_r}dt=-\frac{Q_1+Q_2 e^{-2\phi}}{4H\rho_r}\frac{dz}{(1+z)}
\label{1-22}\end{equation}
where in the second equality we have used $a=(1+z)^{-1}$\footnote{We have normalized the present value of the scale factor to unity, namely that $a_0=1$.}. One can put the interaction functions from (\ref{1-10a}) and (\ref{1-10b}) into (\ref{1-22}) to get
\begin{equation}
\ln\tau(z)=\tau_1(z)+\tau_2(z)
\label{1-23}\end{equation}
where
\begin{equation}
\tau_1(z)=-\frac{1}{2}\int\frac{\dot{\phi}}{H}\frac{dz}{(1+z)}
\label{1-24}\end{equation}
\begin{equation}
\tau_2(z)=-\frac{1}{2}|\zeta|\int\frac{\rho_m}{\rho_r}\frac{\dot{\phi}}{H} e^{-2\phi}\frac{dz}{(1+z)}
\label{1-25}\end{equation}
For evaluating the integrals in (\ref{1-24}) and (\ref{1-25}), we need $\frac{\dot{\phi}}{H}$
which can be obtained from the solution (\ref{1-12}). It gives
\begin{equation}
\frac{\dot{\phi}}{H}=2\beta~~~~~,~~~~~~~~\beta=\frac{1}{n\lambda}
\label{1-24ab}\end{equation}
Then (\ref{1-24}) leads to
\begin{equation}
\tau_1(z)=-\beta\ln(1+z)
\label{1-26}\end{equation}
In radiation domination (corresponding to $z>3000$) $\frac{\rho_m}{\bar{\rho}_{r}}<<1$, then $\tau_2(z)\approx 0$. In this case $\tau(z)$ is solely given by $\tau_1(z)$ and the temperature evolution law becomes
\begin{equation}
T(z)=T_0 (1+z)^{1-\beta}
\label{1-26a}\end{equation}
This expression implies that for a given redshift $z>>1$, the temperature is slightly different from the prediction of standard cosmology with constant $\alpha$. It should be stressed that (\ref{1-26a}) is similar to (\ref{0-1}) except that the latter is supposed to be valid for $z\sim 1$.\\
On the other hand, in matter domination (corresponding to $0<z<3000$) $\frac{\rho_m}{\bar{\rho}_{r}}>>1$ and thus both $\tau_1(z)$ and $\tau_2(z)$ should be taken as contributors to $\tau(z)$.
For evaluating $\tau_2(z)$, we need the ratio of energy densities which is
\begin{equation}
\frac{\rho_m}{\rho_r}=\frac{\rho_m}{\bar{\rho_r}}e^{-2\phi}=\frac{\rho_{0m}}{\bar{\rho}_{0r}}~a~ e^{-2\phi}
\label{1-28}\end{equation}
From (\ref{1-24ab}), one obtains $\phi=-2\beta\ln(1+z)$ which is equivalent to $e^{-2\phi}=(1+z)^{4\beta}$. Thus (\ref{1-28}) leads to
\begin{equation}
\frac{\rho_m}{\rho_r}=\frac{\rho_{0m}}{\bar{\rho}_{0r}}~(1+z)^{4\beta-1}
\label{1-28a}\end{equation}
The integral of (\ref{1-25}) can now be straightforwardly calculated by using of (\ref{1-24ab}) and (\ref{1-28a}). It gives
\begin{equation}
\tau_2(z)=C(1+z)^{8\beta-1}
\label{1-29a}\end{equation}
with $C=-\frac{\beta}{(8\beta-1)}|\zeta|\frac{\rho_{0m}}{\bar{\rho}_{0r}}$ being a constant. Putting (\ref{1-26}) and (\ref{1-29a}) into (\ref{1-23}), we get the final expression
\begin{equation}
\tau(z)=(1+z)^{-\beta}e^{\tau_2(z)}
\label{1-31}\end{equation}
\begin{figure}[h]
\begin{center}
\includegraphics[width=0.45\linewidth]{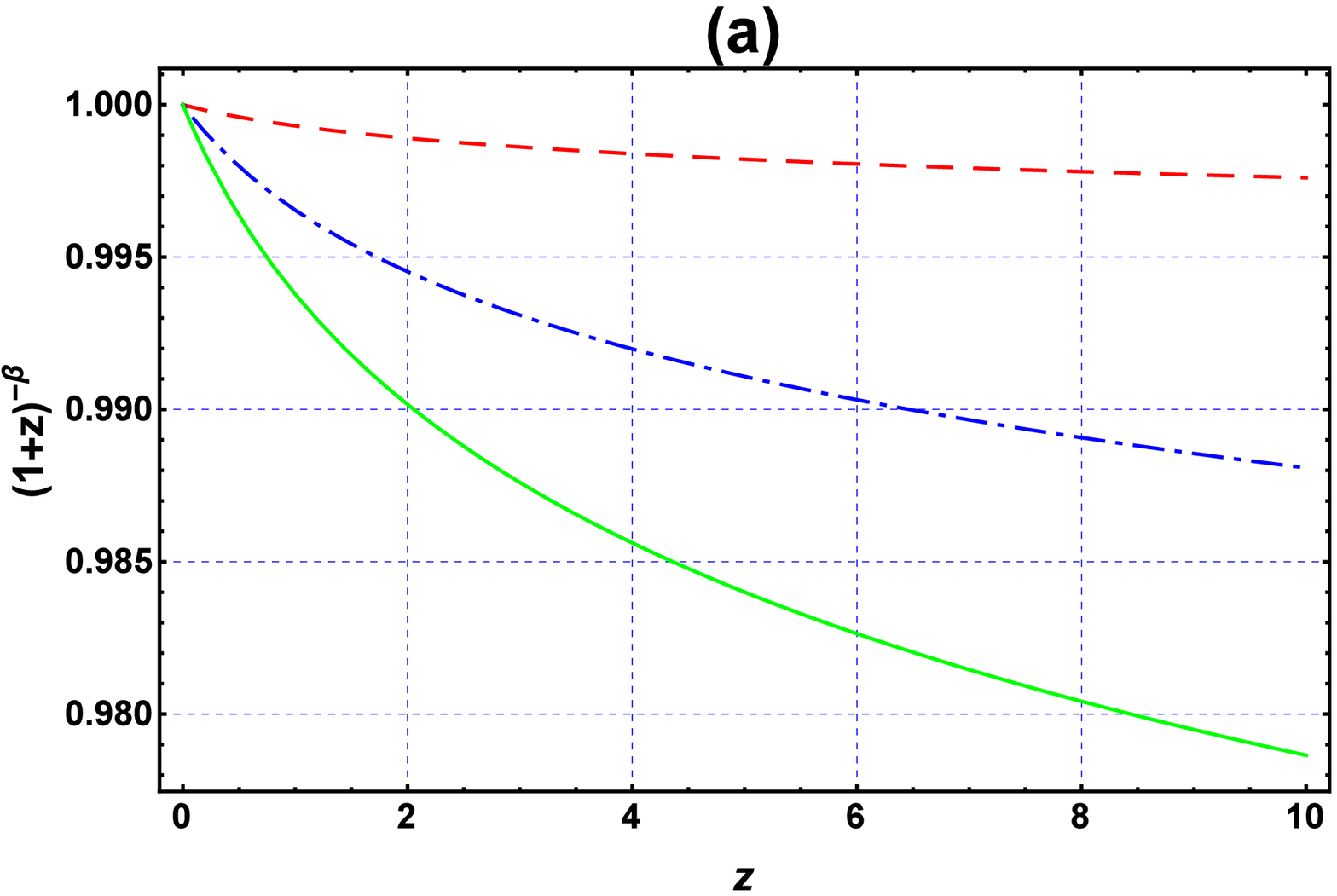}
\includegraphics[width=0.45\linewidth]{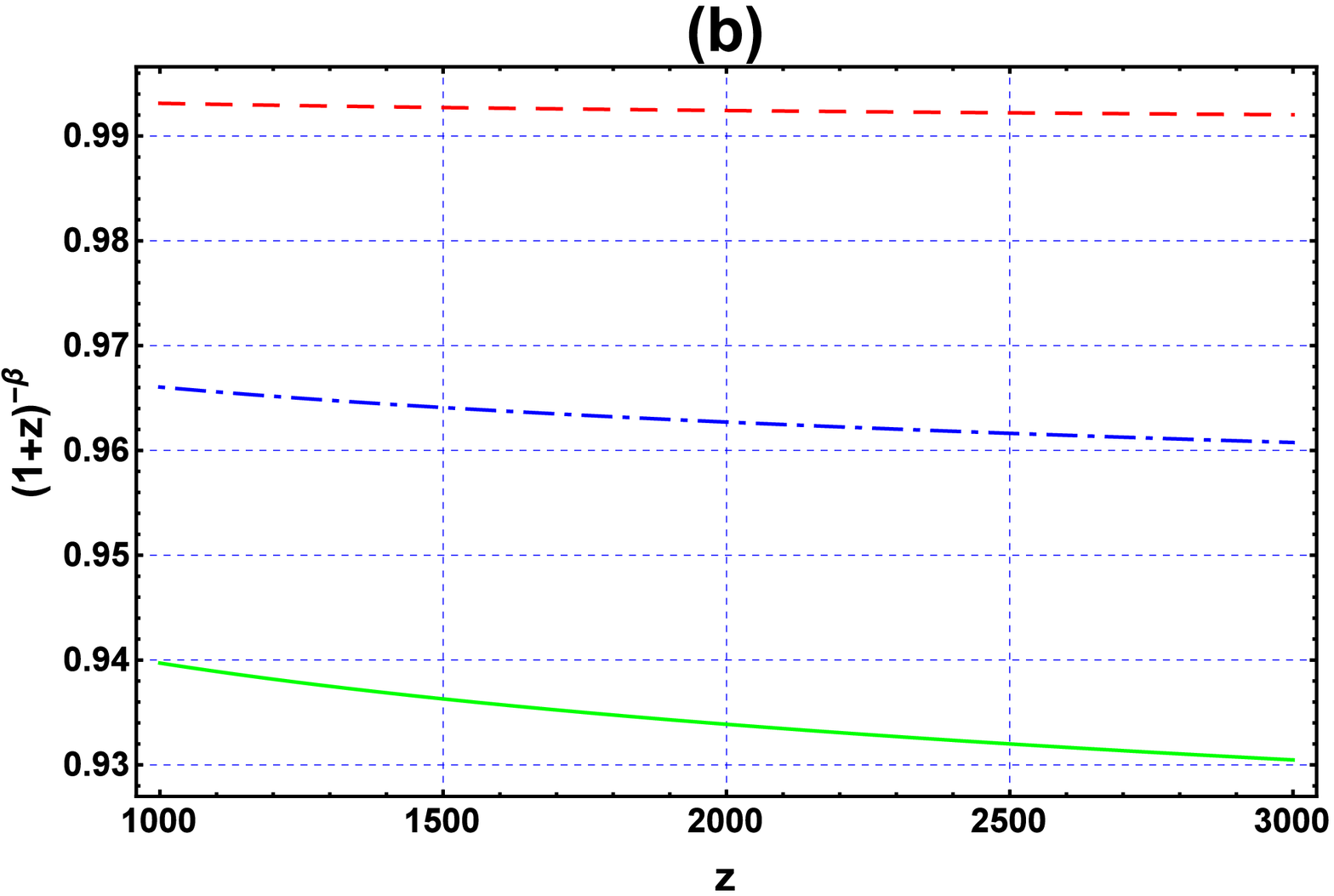}
\end{center}
\end{figure}
\begin{figure}[h]
\begin{center}
\includegraphics[width=0.45\linewidth]{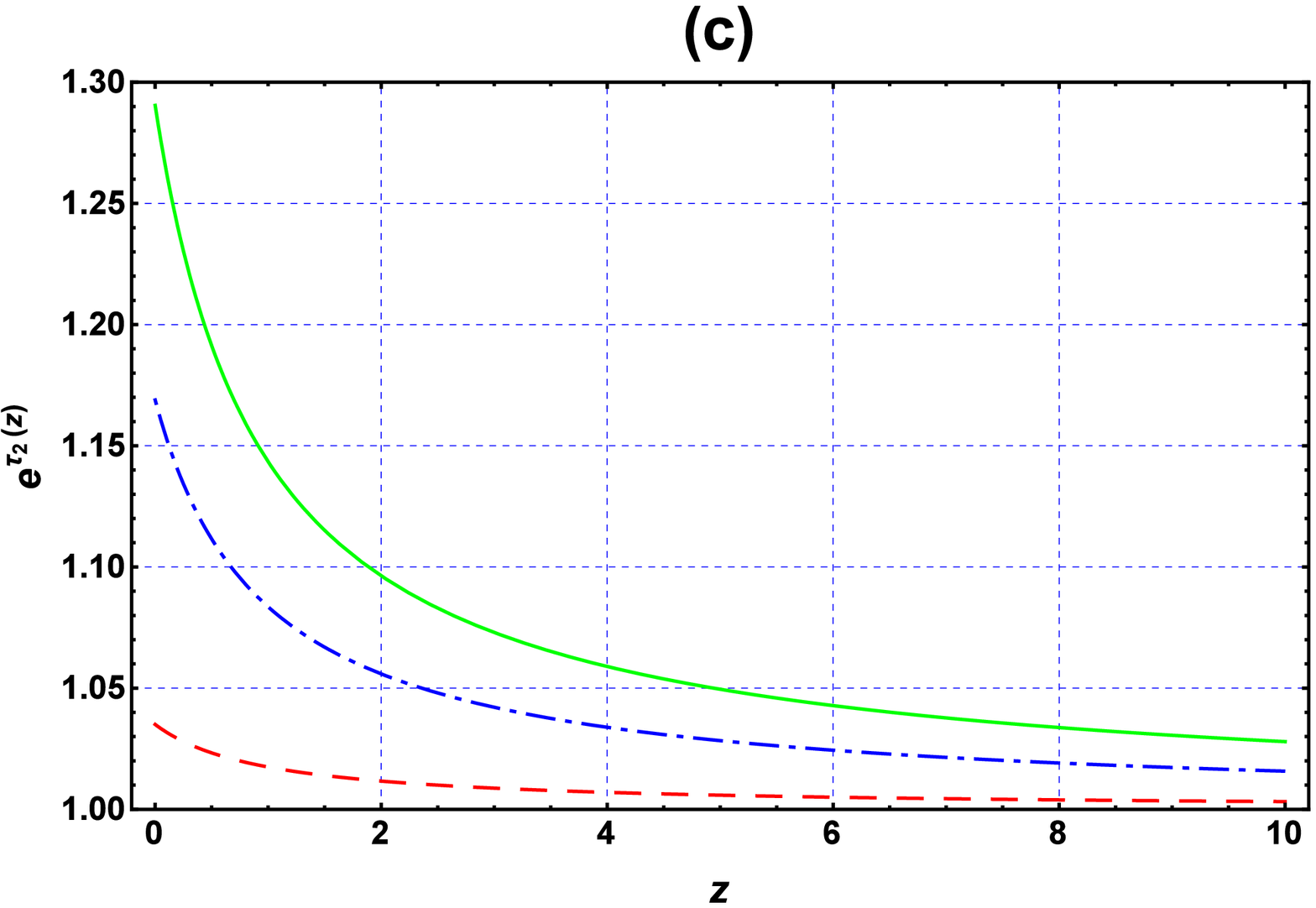}
\includegraphics[width=0.45\linewidth]{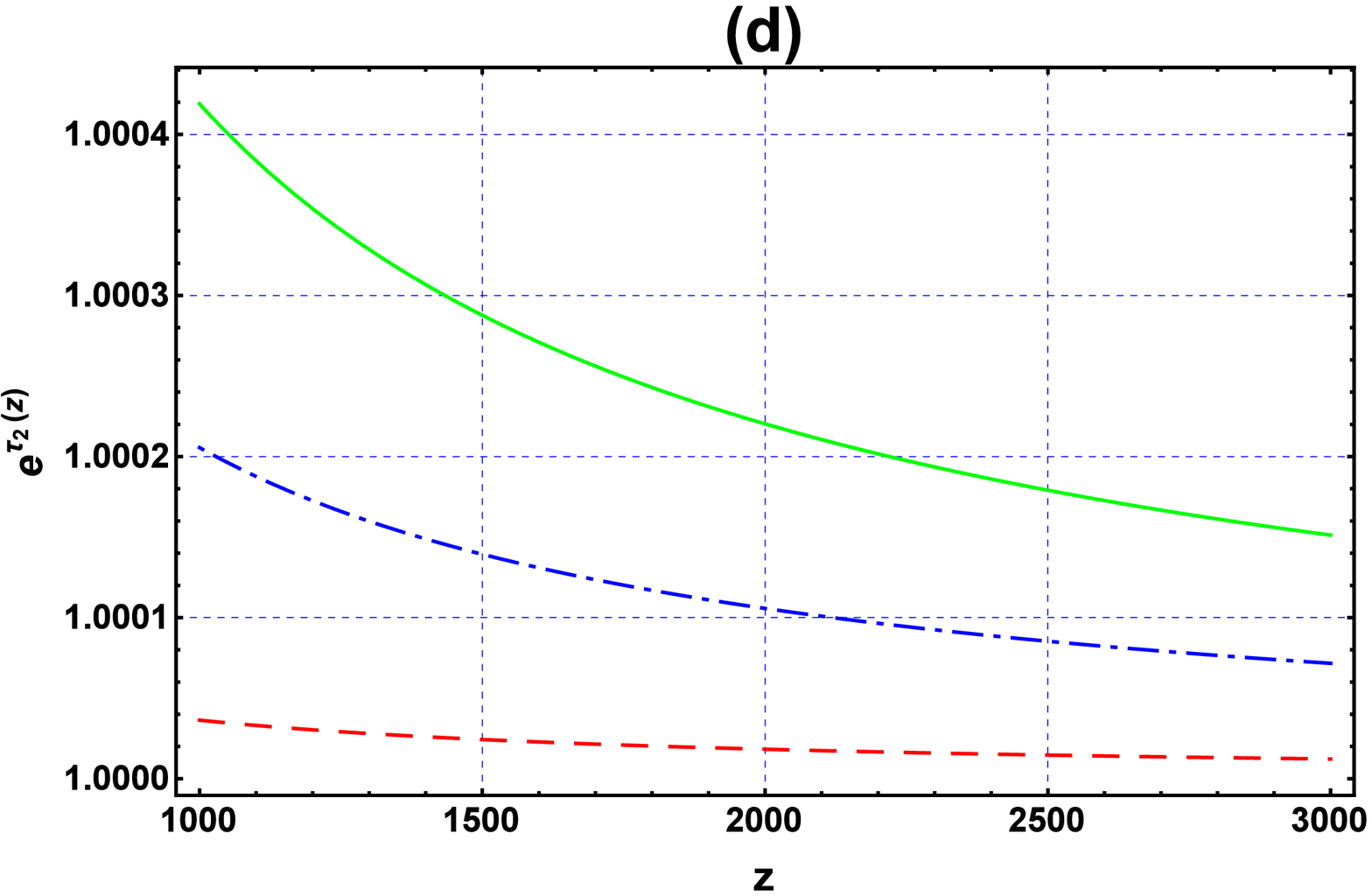}
\caption{The plots of $(1+z)^{-\beta}$ and $e^{\tau_2(z)}$ for $\beta=0.001$ (the dashed lines), $0.005$ (the dashdotted lines) and $0.009$ (the solid lines). We also take $|\zeta|= 0.01\%$, $\Omega_{0m}=0.3$ and $h=0.7$.}
\end{center}
\end{figure}
There is a link between variations of $\alpha$ and the distortion of temperature-redshift relation \cite{link}. The constraint on $\lambda$, obtained in the previous section, gives the bound $\beta\sim 10^{-6}$ implying that $8\beta-1<0$ by which one deduces that when $z\rightarrow \infty$, the coefficient $e^{\tau_2(z)}$ goes to unity and then (\ref{1-21}) reduces to (\ref{1-26a}). This behavior is consistent with the temperature-redshift law (\ref{1-26a}) which was obtained in a radiation-dominated era $\rho_m<<\rho_r$.
One therefore expects that (\ref{1-31}) gives an effective deviation from (\ref{1-26a}) only in the matter-dominated era and $z\rightarrow 0$.
This is indicated in the fig.1. The amount of this deviation depends on
the numerical value of the ratio $\frac{\rho_{0m}}{\bar{\rho}_{0r}}$. To estimate the latter, we use (\ref{1-28a})
to write
\begin{equation}
\frac{\rho_{0m}}{\bar{\rho}_{0r}}=a^{4\beta-1}_{eq}=(1+z_{eq})^{1-4\beta}
\label{1-29}\end{equation}
where $a_{eq}$ and $z_{eq}$ are the scale factor and redshift at the epoch of matter-radiation equality. This epoch is expressed in terms of the redshift as \cite{dod}
\begin{equation}
(1+z_{eq})=2.4\times 10^{4}\Omega_{0m} h^2
\label{1-30}\end{equation}
in which $\Omega_{0m}$ is the density parameter of matter today and $h$ is the parameter characterized by the Hubble constant $H_0=100h~Km~Sec^{-1}~Mpc^{-1}$.
The contribution values of the coefficients $(1+z)^{-\beta}$ and $e^{\tau_2(z)}$ in $T(z)$ is plotted in the fig.1.
It indicates that at early times ($z\rightarrow\infty$), the coefficient $(1+z)^{-\beta}$ slightly modifies $T(z)$ while $e^{\tau_2(z)}\rightarrow 1$. At late times ($z\rightarrow 0$), on the other hand, $(1+z)^{-\beta}$ goes to unity while $e^{\tau_2(z)}$ comes into play.

~~~~~~~~~~~~~~~~~~~~~~~~~~~~~~~~~~~~~~~~~~~~~~~~~~~~~~~~~~~~~~~~~~~~~~~~~~~~~~~~~~~~~~~~~~~~~~~~~~~~~~~~~~~
\section{Discussion}
In this paper, we have investigated redshift dependence of CMB temperature in the BSBM model. This work is divided into two main parts. In part one, the BSBM model is reviewed in the case that the scalar field $\phi$ is allowed to have a self-interacting exponential potential. The solution of the field equations indicate that $\phi$ increases logarithmically both in the radiation and the matter dominated eras.
The slope of the evolution of $\phi$ is characterized by the parameter $\lambda$. The larger $\lambda$ is, the smaller is the rate of change of $\phi$ and also the less important is the potential term. This parameter is then constrained by recent tight bounds on temporal variation of $\alpha$. These observational bounds require that $\lambda\sim 10^{6}$ which implies that any changes of $\alpha$ would be very tiny.\\
In the second part, we explore deviation from the standard evolution of CMB temperature. In the adiabatic approximation, we have shown that such a deviation consists of two components $\tau_1(z)$ and $\tau_2(z)$ generated by $Q_1$ and $Q_2$ respectively. Our analysis allows us to establish an analytic relation between:\\
a) the source term $Q_1$ and the expression (\ref{1-26a}) at high redshifts eras although, such a parametrization is usually used to measure deviations from the standard redshift-temperature law by low redshift observations \cite{non}.\\
b) the source term $Q_2$ and the expressions (\ref{1-21}) and (\ref{1-31}). Although $Q_2$ may seem that has not an important role in the field equations and may be neglected, it could affect the CMB temperature by the factor $e^{\tau_2(z)}$ when $z\rightarrow 0$. Therefore, (\ref{1-21}) and (\ref{1-31}) give deviations from the standard redshift-temperature law at low redshifts and should be taken as a basis for constraining $\beta$ at $z\sim 1$.\\
We have found that $\beta\propto \lambda^{-1}$ which establishes a link between two different kinds of observations. In one hand, observations of $z$-dependence of temperature of CMB which leads to constraints on $\beta$ and, on the other hand, observations of $\alpha$-variation which leads to constraints on $\lambda$. The latter observations may be also regarded as a basis for constraining the parameter $\beta$.\\
The main result of this work is enhancement of the effect of $\tau_2(z)$ at late times and gradual departure of the redshift-temperature law from the ansatz (\ref{1-26a}). This may have important impacts on:\\
a) CMB anisotropy spectrum so that the data given by COBE and PLANCK or future missions may need to be reinterpreted by new parameterizations.\\
b) future observations of 21 cm emission from neutral hydrogen. These signals probe the cosmic
"Dark Ages" comprising the epoch between the last scattering of CMB at $z\sim 1100$ and the appearance of luminous sources at
$z\sim 30$ \cite{21a}. It is shown that the $21$ cm signal is affected by possible $\alpha$-variations \cite{21b} but the relevance of $Q_1$ and $Q_2$ on its temperature and wavelength scaling laws can be explored by future surveys, such as those done by square kilometre array (SKA).\\

{\bf Acknowledgment}\\\\
The author is acknowledged the anonymous referees for useful comments. This work is supported by the Iran National Science Foundation (INSF) with Grant No.6074.

~~~~~~~~~~~~~~~~~~~~~~~~~~~~~~~~~~~~~~~~~~~~~~~~~~~~~~~~~~~~~~~~~~~~~~~~~~~~~~~~~~~~~~~~~~~~~~~~~~~~~~~~~~~

\end{document}